\documentstyle[12pt]{article}
\input epsf
\begin{document}

\begin{titlepage}
\begin{tabular}{l}
\end{tabular}
    \hfill
\begin{tabular}{l}
hep-ph/9801444 \\
June 17, 1998 \\
\end{tabular}

\vspace{2cm}

\begin{center}

\renewcommand{\thefootnote}{\fnsymbol{footnote}}

{\LARGE Study of the Uncertainty of the Gluon Distribution\footnote[1]{
This work was supported in part by the DOE and NSF.}}

\renewcommand{\thefootnote}{\arabic{footnote}}

\vspace{1.25cm}


{\large J.~Huston$^d$, S.~Kuhlmann$^a$, H.~L.~Lai$^{d,e}$, 
F. Olness$^{b,g}$, \\ J.~F.~Owens$^c$, D.~E.~Soper$^f$, W.~K.~Tung$^d$}

\vspace{1.25cm}

$^a$Argonne National Laboratory, 
$^b$Fermi National Accelerator Laboratory, \\
$^c$Florida State University,
$^d$Michigan State University, \\ 
$^e$National Tsing Hua University,
$^f$University of Oregon, \\ 
$^g$Southern Methodist University

\end{center}
\vfill

\begin{abstract}
The uncertainty in the calculation of many important
new processes at the Tevatron and LHC is dominated by that
concerning the gluon distribution function.
We investigate the uncertainty in the gluon
distribution of the proton by systematically varying the gluon
parameters in the global QCD analysis of parton distributions.
The results depend critically on the parton momentum fraction $x$ and 
the QCD scale $Q^2$.
The uncertainties are presented 
for integrated gluon-gluon and gluon-quark luminosities for both
the Tevatron and LHC as a function of 
$\sqrt{\tau}=\sqrt{x_1 x_2}=\sqrt{\hat{s} /s}$, the most 
relevant quantity for new particle production.
The uncertainties are reasonably small, except for 
large $x$.  
\end{abstract}

\vfill
\newpage
\end{titlepage}

\renewcommand{\baselinestretch}{1.25}

Many hadron-collider signatures of physics beyond the Standard Model 
have a gluon in the initial state,  
in either the signal process or the important background 
processes.  
Other new signatures within the Standard Model can also
have gluons in the initial state.  
One example of this is the production of a light Higgs boson 
at the LHC via the 
process~$gg\rightarrow H \rightarrow \gamma \gamma$.
Another example is the measurement of $V_{tb}$ using single-top 
production at the Tevatron via the process $ gW\rightarrow tb $.
It is important to estimate the theoretical uncertainty in 
the Quantum Chromodynamics (QCD)~calculations of these 
new processes.  
Since the quark distributions of the nucleon
are relatively well-determined,
the dominant uncertainty in these cases is due to that of the
gluon distribution.  
The conventional method of estimating parton 
distribution uncertainties is to compare different 
published parton parameterizations.  This is a completely 
unreliable approach since the authors of most published sets of 
parton distributions adopt similar assumptions and use
similar data sets.  
The differences between these sets have little
to do with the range of possible variations
of the parton distributions as constrained by current theory and
available data.  
In this paper we focus on the uncertainty of the 
gluon distribution within the framework of 
the CTEQ global QCD analysis~\cite{cteq4}, and present a more complete 
estimate of the uncertainties.  Not surprisingly, 
we will find that the uncertainty is a function of the 
gluon $x$ and $Q^2$. 

Ideally, one might hope to perform a full error analysis
and provide an error-correlation matrix for all the parton distributions.
This ambitious goal is, however, impractical at this time for two 
reasons.  First, only a subset of available experiments
provide correlation information for their data sets in a way 
suitable for this analysis.
Secondly, there is no established way
to quantify the theoretical uncertainties
for the diverse physical processes used in the global analysis.
One possibility,
explored in reference~\cite{alekhin}, is to invoke only the 
deep-inelastic scattering (DIS) process, to use the DIS 
data sets with the needed correlation information,
and only use those data points at high $Q^2$ where
the theoretical uncertainties are expected to be negligible.
While this procedure is of methodological interest,  it 
leaves out many useful data sets 
and the uncertainties obtained for the gluon are clearly unrealistic.

The approach we adopt in this paper is to systematically vary
the gluon distribution parameters in the 
global analysis framework. 
We then conservatively delineate the range of admissible
distributions as that bounded by fits which show clear
disagreements with more than one data set.  For this purpose,
we adopt the CTEQ4M parton distribution set~\cite{cteq4} 
as the standard and explore the range of possible variations of the gluon
distribution around it.
The conclusions of this study should apply to all modern
parton distribution sets since they are in rather
good agreement with each other.~\cite{cteq4,mrs}

\paragraph{Constraints Based on the Momentum Sum Rule}
The momentum fraction of the proton carried by quarks is determined
by deep-inelastic scattering data to be 58\% in the CTEQ4M 
analysis (Q=1.6 GeV)~\cite{cteq4}.
The uncertainty in this number is mainly due to normalization 
uncertainties of the experimental data sets,  which is 
typically $\pm 2\%$.  Therefore the total gluon momentum fraction 
in the CTEQ4M fit is 42\% with an uncertainty of about 2\%.  
This is an extremely important constraint that is not 
fully appreciated.  If the flux of gluons in a 
certain $x$ range is increased, the flux must be reduced
by almost the same amount somewhere else.

\begin{table}[tbph]
\begin{center}
\begin{tabular}{|c|c|}\hline
x Bin & Momentum Fraction \\ \hline
$10^{-4}$ to $10^{-3}$ & 0.6\% \\
$10^{-3}$ to 0.01 & 3\% \\
0.01 to 0.1 & 16\% \\
0.1 to 0.2 & 10\% \\
0.2 to 0.3 & 6\% \\
0.3 to 0.5 & 5\% \\
0.5 to 1.0 & 1\% \\ \hline
\end{tabular}
\end{center}
\caption{The fraction of proton momentum carried by gluons in 
different $x$ bins.
This is for the CTEQ4M parameterization at Q=1.6 GeV.}
\label{tab1}
\end{table}

\begin{figure}[tbph]
\begin{center}
\begin{minipage}[h]{6.5in}
\epsfxsize=6.3in
\epsfbox[36 144 520 650]{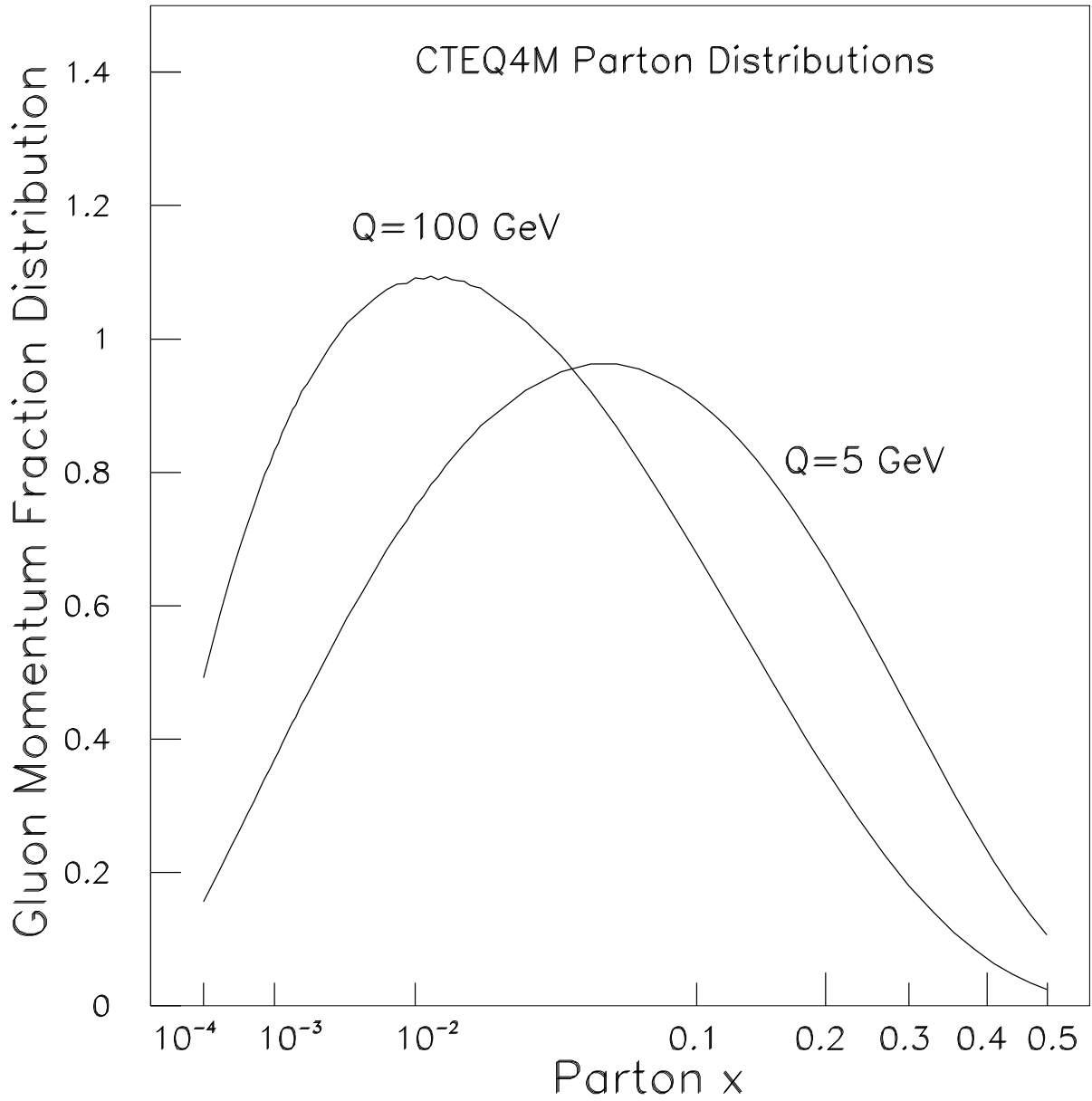}
\caption{The gluon momentum fraction distribution is shown for 
Q=5 GeV and for Q=100 GeV.}
\label{glumom}
\end{minipage}
\end{center}
\end{figure}

Table~\ref{tab1} shows how the momentum fraction of gluons 
within the proton is distributed as a function of $x$ 
for the CTEQ4M parameterization at Q=1.6 GeV.  
The largest component of the gluon momentum is carried at 
medium values of $x$, 
since this has the largest product of the number of 
gluons and the momentum fraction of each gluon.  
A simple exercise using these momentum fractions 
is illustrative.  If we assume the flux of gluons 
in the range $0.01<x<0.3$ is decreased by 20\% below the CTEQ4M 
value, what happens to the momentum sum rule constraints?
From Table~\ref{tab1}, 32\% of the proton momentum is
in this $x$ region, a 20\% decrease is 6.4\% which is the 
amount that has to be increased elsewhere.  The compensation 
would have to come from a $\times 2.8$ increase ((6.4+3.6)/3.6) 
in gluons below $x=0.01$,  or a $\times 2.1$ increase ((6.4+6)/6)
in gluons above $x=0.3$, or a combination of the two.
Typical uncertainties from HERA (at Q=1.6 GeV) in the 
gluon distribution for small $x$ are shown to be 30-40\%~\cite{dis97}, 
therefore not much compensation
can come from small $x$.  At larger $x$ the fixed 
target Drell-Yan data is sensitive to the gluon 
since the sea quarks couple to the gluons.
As shown in the next section, the increase in the gluon distribution 
described above would increase the predicted Drell-Yan cross 
sections at large $x$ by $>40\%$,  ruining the present agreement with 
CTEQ4M parton distributions.
Therefore the needed compensation for the 20\% change in gluons is unlikely
to come from $x>0.3$ either.   This exercise illustrates 
the important constraint on the gluon distribution at 
medium $x$ from the momentum sum rule,  and also serves 
to explain the quantitative results on parton distribution
uncertainties discussed 
in the following sections.  Naturally this exercise is 
simplified since the momentum fraction of the gluon 
changes with $Q^2$.  In Fig.~\ref{glumom} we plot the 
gluon momentum fraction distribution for Q=5 GeV and for 
Q=100 GeV.  In this plot the area under the curve 
in any $x$ interval is the gluon momentum fraction in that
region.  The evolution to smaller parton $x$ as Q increases
is evident,  but in both cases the bulk of the gluon 
momentum is at medium values of $x$. 

\paragraph{Scanning the Gluon Parameters }
We now perform a detailed study of the range of possible variation
of the gluon distribution by systematically varying the gluon parameters
in a global analysis.  The CTEQ4 gluon parametrization
is: $A_0 x^{A_1}(1-x)^{A_2}(1+{A_3}x^{A_4})$, with
$A_0=1.1229,\ A_1=-0.206,\ A_2=4.673,\ A_3=4.269,\ A_4=1.508$
for the standard CTEQ4M parton distribution set.
We begin by fixing
$\alpha_S$ to be the CTEQ4M value($\alpha_S(M_Z)=0.116$), 
which is close to the current world average of 
$0.118 \pm 0.003$.~\cite{alpha}
We have studied the variation of the gluon with $\alpha_S$ 
in a previous publication~\cite{cteq4}, and will discuss
this more later.
We have then systematically varied the values of 
$A_1, A_2,$ and $A_3$, each time refitting 
the other quark and gluon parameters using the CTEQ 
procedure described in Ref.~\cite{cteq4}.\footnote{
We found the variation of $A_3$ was 
easily compensated by changes in $A_4$, and vice versa.  
Therefore in the $A_3$ parameter scan described below $A_4$
was fixed to the CTEQ4M value.}
The range of variation of each parameter was expanded until
clear disagreements with more than one data set were observed.
In order to be conservative in this study, we performed these scans 
using only the well-established DIS and Drell-Yan data sets. This
also allows us to establish a baseline uncertainty estimate 
with the processes that are best understood theoretically. 
We will discuss the possible impact of direct photon and jet production
data in a later section.

\begin{figure}[tbph]
\begin{center}
\begin{minipage}[h]{6.5in}
\epsfxsize=6.3in
\epsfbox[36 144 520 650]{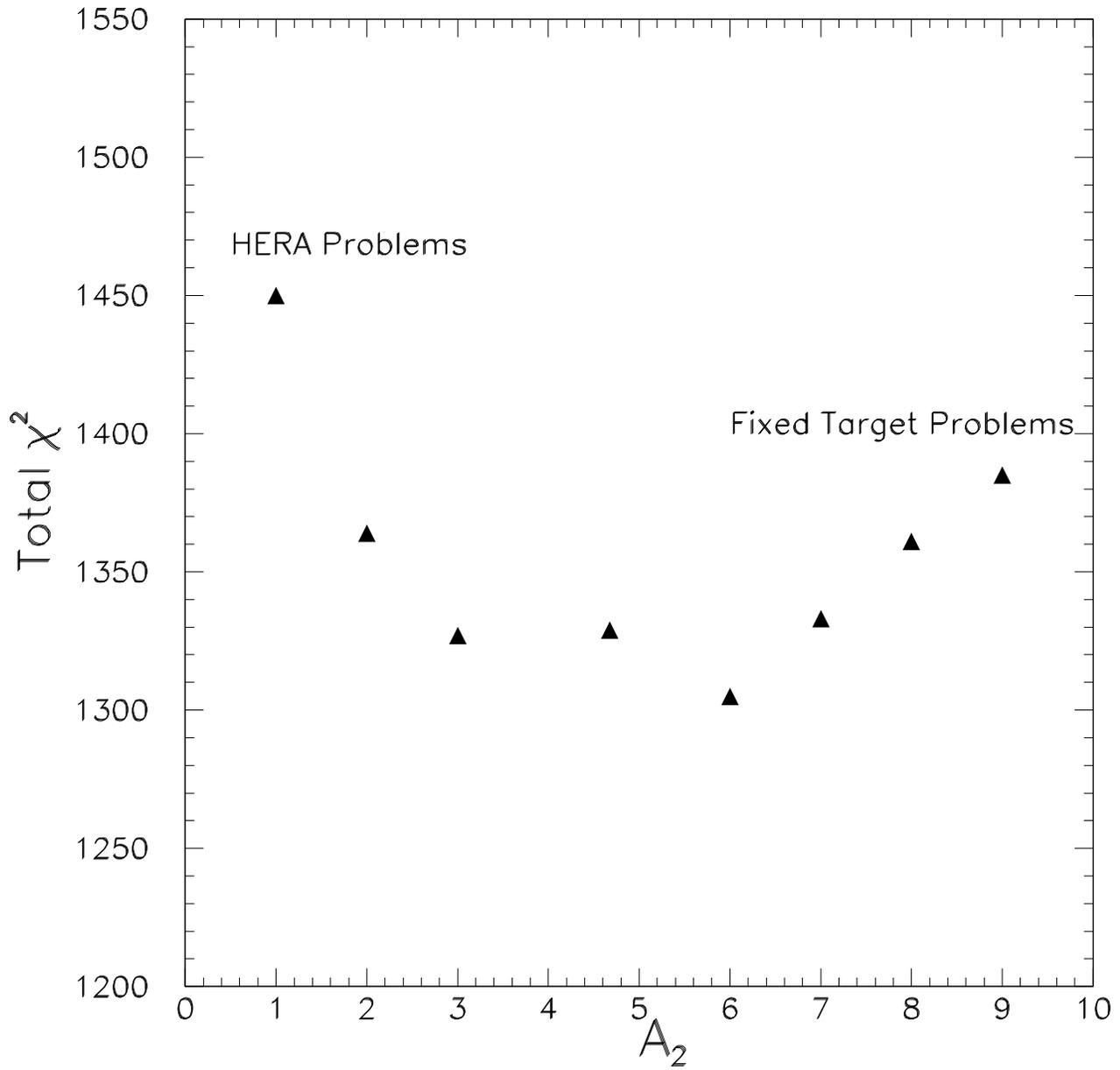}
\caption{The total $\chi^2$ is shown from the $A_2$ parameter
scan is shown.}
\label{chisq}
\end{minipage}
\end{center}
\end{figure}

\begin{figure}[tbph]
\begin{center}
\begin{minipage}[h]{6.5in}
\epsfxsize=6.3in
\epsfbox[36 144 520 650]{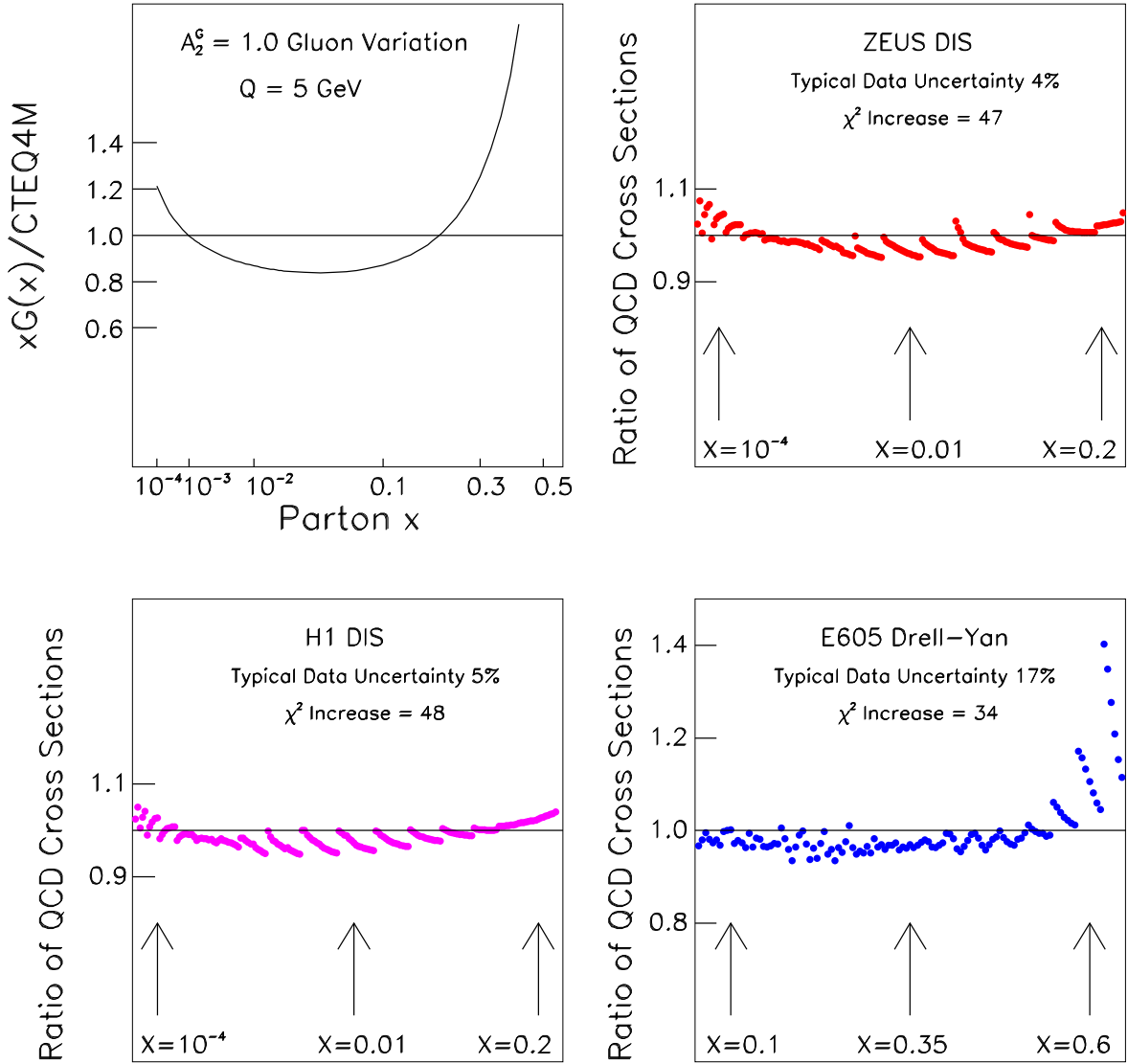}
\caption{One example of a gluon distribution that 
causes clear disagreements with data is shown. 
Upper left is the ratio of the gluon distribution 
to CTEQ4M.  The other three plots show the ratio 
of QCD predictions for three of the data sets, 
using the trial gluon distribution.  Also 
indicated on these three plots are the typical data 
uncertainty, and the change in $\chi^2$ for 
this set of data.}
\label{foursli}
\end{minipage}
\end{center}
\end{figure}

\begin{figure}[tbph]
\begin{center}
\begin{minipage}[h]{6.5in}
\epsfxsize=6.3in
\epsfbox[36 144 520 650]{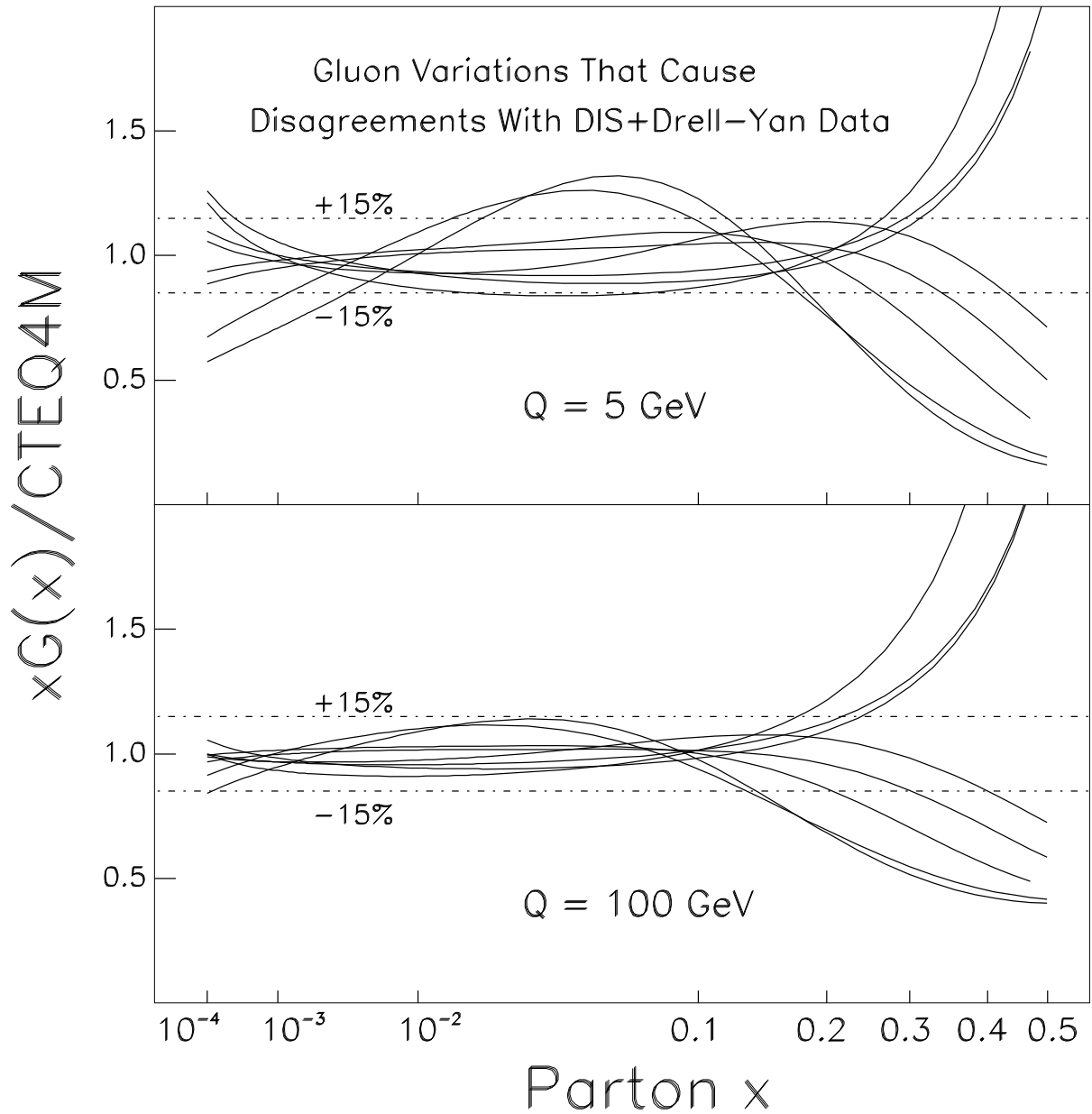}
\caption{The ratio of gluon distributions compared to CTEQ4M 
is shown.  On top is for Q=5 GeV,  and on bottom is Q=100 GeV. 
These are the examples that cause clear disagreements with 
some DIS+Drell-Yan data sets (see text).}
\label{allbad}
\end{minipage}
\end{center}
\end{figure}

As an example of one of these parameter scans,  the 
total $\chi^2$ from the $A_2$ scan 
is shown in Fig.~\ref{chisq}.  
The parameter $A_2$ is the exponent of the $(1-x)$ 
factor.  
It is varied over the wide range from 1.0 to 9.0, 
with the CTEQ4M value being 4.673. 
The total $\chi^2$ variable was only used for guidance
in determining acceptable gluon distributions, 
a strict cut was not applied.
In practice we examined closely every data set 
for every variation, and coupled with our 
experience with experimental and theoretical 
uncertainties, determined which gluon 
distributions caused disagreements with 
data that could not be explained 
by such uncertainties.
The four worst fits in Fig.~\ref{chisq} have examples of 
clear disagreements with some data sets.
Figure~\ref{foursli} shows the 
change in the gluon distribution, and the 
corresponding effect on three data sets for the 
$A_2=1.0$ fit.  This change in gluon distribution
is almost exactly that described earlier 
concerning the momentum sum rule, the upper 
left figure demonstrates this.  
In another example, the $A_2=9.0$ variation, 
all of the fixed target DIS data sets 
(with 60-170 data points each) had increased $\chi^2$ of 
15 units or more over those of the CTEQ4M fit.
Similar criteria were applied for extreme fits 
to the $A_1, A_3$ parameter scans.  Fig.~\ref{allbad} 
shows the ratio of the gluon distributions from 
these extreme fits to CTEQ4M, 
with Q=5 GeV on top and Q=100 GeV on bottom. 
The effect of the momentum sum rule constraint 
is once again dramatically demonstrated at Q=5 GeV.  
The relatively small changes at moderate values
of $x$ are compensated by large changes at small and 
large $x$.  
The range of gluon distributions at Q=100 GeV is
interesting since the smallest invariant mass
for typically new particle production
is around M=100 GeV.  Notice that the variation of the gluon is only
$\pm 15\%$ below $x=0.15$ at this scale; and these 
are the fits which already show clear conflicts with 
existing data sets, some of which cause increases 
in $\chi^2$ of more than 200 units.  
This demonstrates how QCD evolution tends to wash out 
the differences in parton distributions at low $Q$.  At larger
$x$ the differences between the gluon variations 
remain fairly constant as $Q$ increases.  This is 
because of the influence of even larger differences
in the gluon distribution for $x>0.5$ (not
shown on the plot) that are feeding down into the 
$x=0.2-0.5$ region at higher $Q^2$.

\begin{figure}[tbph]
\begin{center}
\begin{minipage}[h]{6.5in}
\epsfxsize=6.3in
\epsfbox[36 144 520 650]{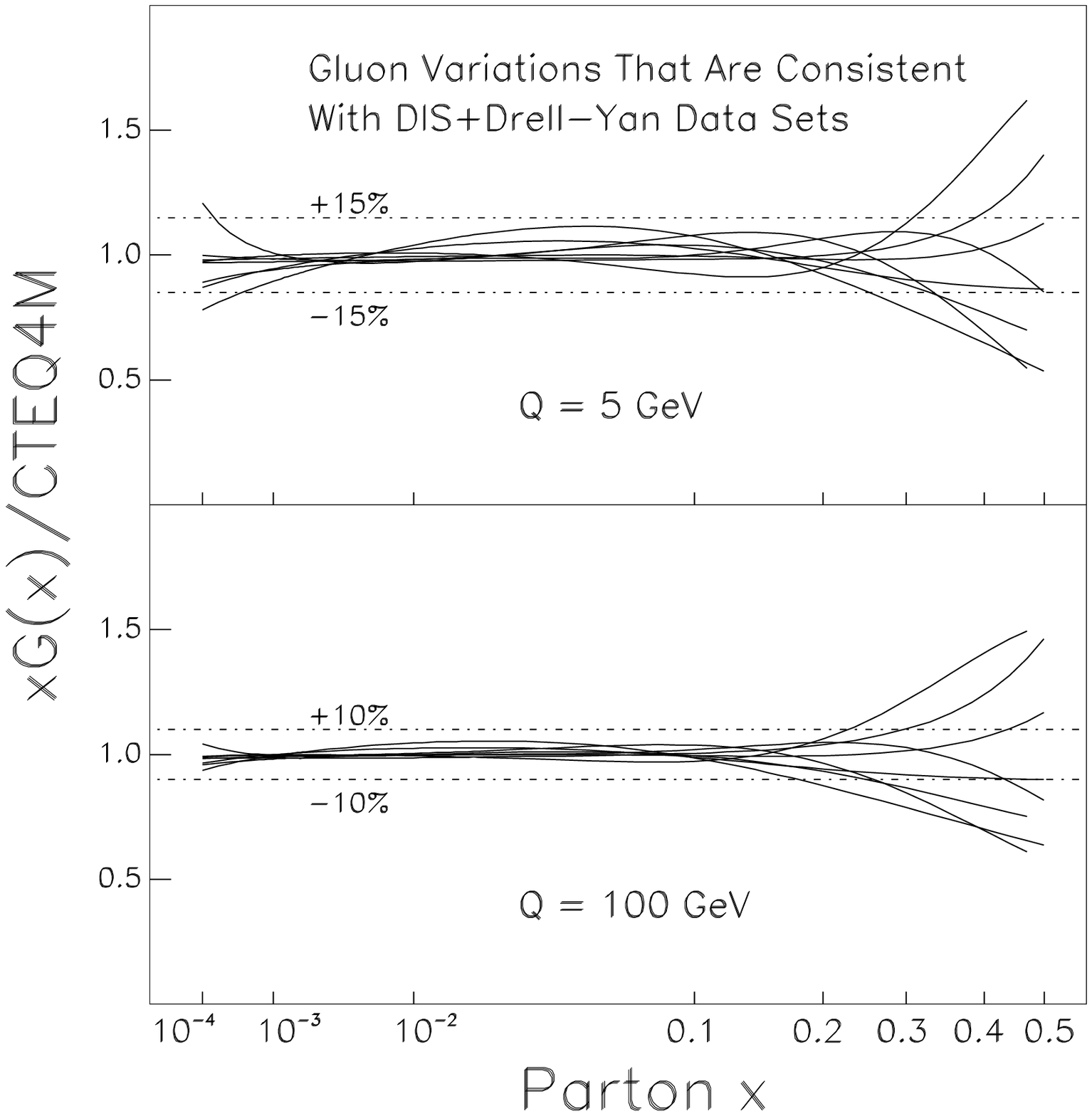}
\caption{The ratio of gluon distributions compared to CTEQ4M 
is shown.  On top is for Q=5 GeV,  and on bottom is Q=100 GeV. 
These are the examples that are consistent with 
DIS+Drell-Yan data sets (see text).}
\label{allgood}
\end{minipage}
\end{center}
\end{figure}

A reasonable estimate of the current uncertainties on the gluon
distribution can be obtained by examining the range of
variations spanned by those fits which do not clearly contradict any of
the data sets used.  Fig.~\ref{allgood} shows the result
obtained from all such scans.
The pattern seen is similar to that shown in Fig.~\ref{allbad}. At
$Q=100$ GeV, the range of variation for the gluon
is relatively small below $x<0.15$;
it increases steadily as $x$ increases.  
Below $x<0.15$ the
range of gluons appears to be of the order 7\%.
This may not yet be the true range of
gluon distributions since a fixed value of $\alpha_S$ has been used
in this study, whereas $\alpha_S$ and $G(x)$ are known to be
correlated in the global analysis.
It is useful to decouple the two effects since $\alpha_S$ can be
measured in a variety of ways independent of parton distributions, and
these measurements are continuing to improve.
At present the PDG value of $\alpha_S$ is $0.118 \pm 0.003$,  a 
2.5\% uncertainty.   We have varied $\alpha_S$ by 8\%, between 0.113 and 
0.122, and found a 3\% variation in the gluon distribution 
below $x<0.15$ at Q=100 GeV, this sets the scale for 
the additional uncertainty in the gluon distribution 
due to $\alpha_S$ variations.
In addition,  we have taken the correlation between 
$\alpha_S$ and the gluon distribution into account by 
refitting the previous extreme variations 
that caused conflicts with 
present data sets, this time allowing $\alpha_S$ to vary.  
The general conclusions remain the same as that stated
before, with a slight increase in the magnitude of the uncertainty:
the range of gluon distributions is within 10\% of CTEQ4M
below $x<0.15$ and $Q>100$ GeV,  and the uncertainties
grow significantly at larger $x$.

One concern is that the relatively small range of variation on the 
gluon distribution may be an 
artifact of a too-restrictive parameterization, coupled 
with the constraints of the momentum sum rule.
To answer this question, we have modified the gluon distribution 
parameterization according to the following {\it ansatz}:
\begin{equation}
x G(x) = 
x G^{\rm CTEQ4}(x)
+ A_5 (1-x)^5 (x - 1/7).
\end{equation}
The added term affects mainly the shape of the gluon 
distribution at medium $x$.  It is small compared to the 
standard CTEQ4M contribution for $x \ll 1$ and $x \to 1$. 
Furthermore, the total integrated momentum fraction 
contributed by this term is zero, so that the gluon 
distribution is not forced to change at small or large 
$x$ in order to maintain the momentum sum rule.  We 
assessed possibilities for modifying the gluon distribution
at medium $x$ by varying values for $A_5$. 
As in the previous parameter scan, for each value of $A_5$ we 
refit the other gluon and quark parameters.  In this case
we fixed the gluon's $A_3$ and $A_4$ to the CTEQ4M value, 
since these parameters 
have a similar effect at medium $x$ to that of $A_5$.
We then varied $A_5$ until 
clear disagreements with some data sets arose. 
The fits resulting from this study gave rise to gluon distributions
entirely within the bands shown in Fig.~\ref{allgood}.  Thus, we
are confident that the quoted uncertainties are not an 
artifact of the gluon distribution parameterization.

\paragraph{Uncertainty on the gluon-gluon luminosity function}
For assessing the range of predictions on cross-sections for
Standard Model and new physics processes, it is more important
to know the uncertainties on the gluon-gluon and gluon-quark
luminosity functions at the appropriate kinematic region
(in $\tau=x_1 x_2=\hat{s}/s$), rather than on the parton
distributions themselves.\cite{ehlq} 
Therefore, we
turn to the relevant integrated parton-parton luminosity functions.
The gluon-gluon luminosity function is defined to be:
\begin{equation}
\tau dL/d\tau = \int_{\tau}^{1}G(x,Q^2)G(\tau/x,Q^2)dx/x.
\end{equation}
This quantity is directly proportional to the cross-section for the
s-channel production of a single particle; it also gives a good
estimate for more complicated production mechanisms.  
It is most appropriate when the experimental acceptances 
do not play a major role in the cross section calculation. We have 
calculated the gluon-gluon luminosity function
for the parton distribution variations 
shown in Fig.~\ref{allgood}. This was done
for both the Tevatron and LHC energies.

\begin{figure}[tbph]
\begin{center}
\begin{minipage}[h]{6.5in}
\epsfxsize=6.3in
\epsfbox[36 144 520 650]{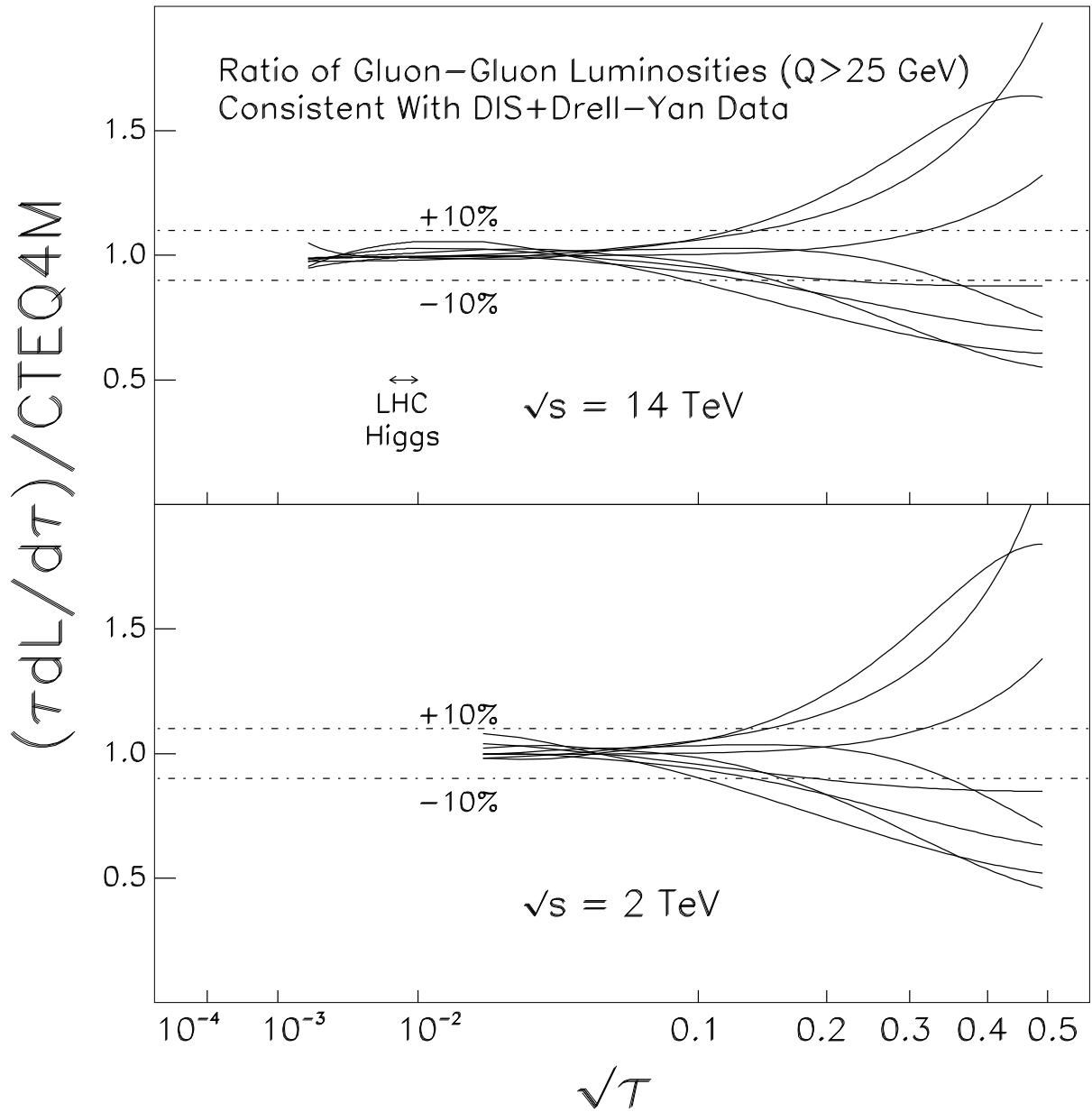}
\caption{The ratio of integrated gluon-gluon luminosities compared
to CTEQ4M is shown as a function of $\sqrt{\tau}$.
These are the examples that are consistent with
DIS+Drell-Yan data sets.}
\label{taugg}
\end{minipage}
\end{center}
\end{figure}

Figure~\ref{taugg} shows the ratio of the  
gluon-gluon luminosity normalized to the 
corresponding result from CTEQ4M, for the variations 
discussed in the last section, as a function of $\sqrt{\tau}$.
Here we take $Q^2 = \tau s$, which naturally takes the 
$Q^2$ dependence of the gluon distribution 
into account as one changes $\sqrt{\tau}$.  This choice
of scale parameter is a common choice in a lowest order 
calculation.
The bands are cutoff below $Q=25$ GeV,  since this region 
is almost certainly irrelevant for new particle production 
at the Tevatron or LHC.  The top figure is for the LHC 
($\sqrt{s}=14$ TeV), and the bottom figure is for the Tevatron 
($\sqrt{s}=2$ TeV).  
The region of production of
a 100-140 GeV Higgs at the LHC is indicated, it lies in the 
medium $x$, large $Q^2$ region where the range of 
variation is $10\%$.  
The size of the bands for $x>0.1$ 
has now grown since we are squaring the variations seen in 
Fig.~\ref{allgood}.  This emphasizes the need for much more 
quality information about the parton distributions at 
large $x$ than is available from DIS+Drell-Yan data sets used
in this analysis.

\begin{figure}[tbph]
\begin{center}
\begin{minipage}[h]{6.5in}
\epsfxsize=6.3in
\epsfbox[36 144 520 650]{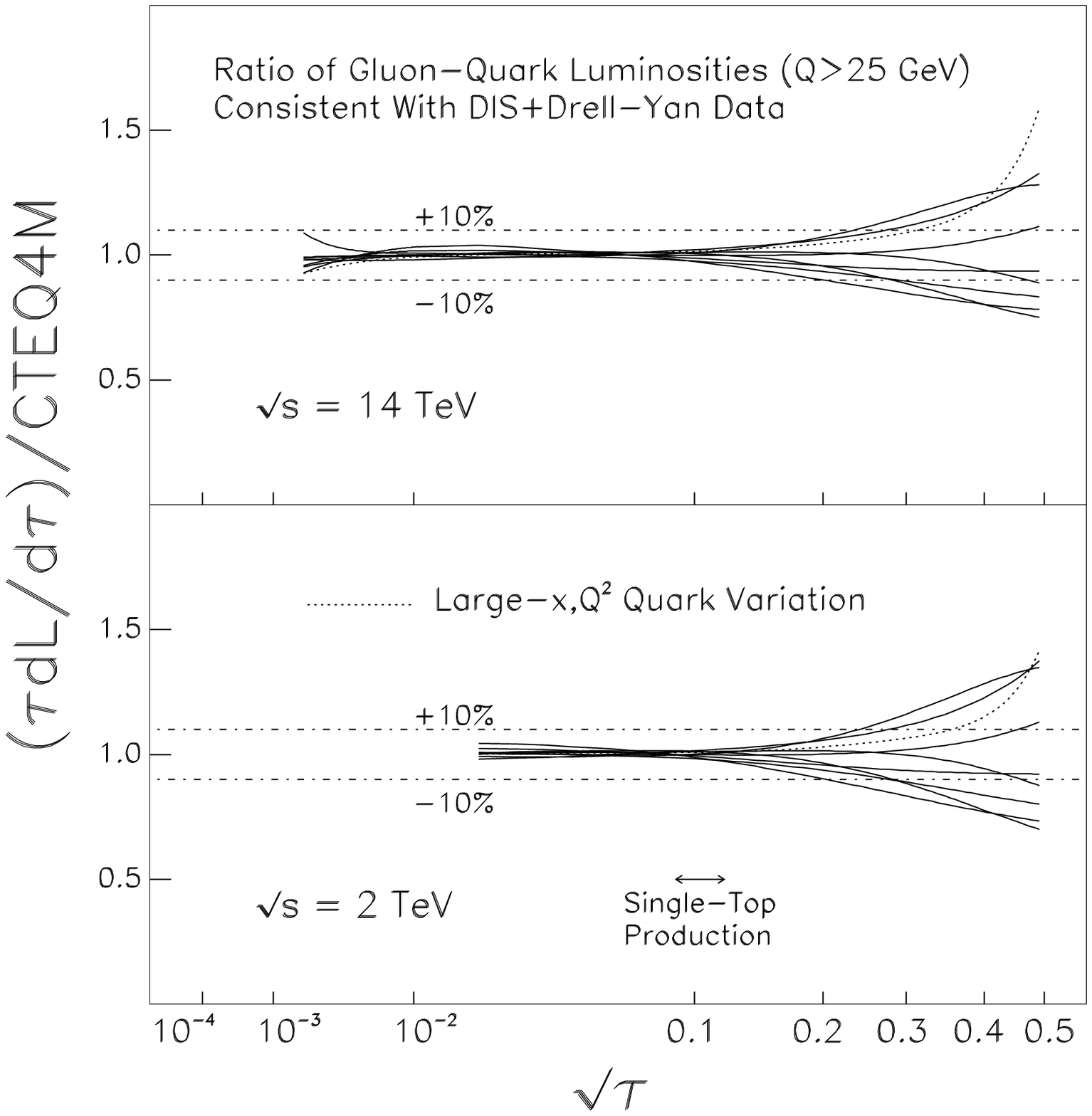}
\caption{The ratio of integrated gluon-quark luminosities compared
to CTEQ4M is shown as a function of $\sqrt{\tau}$.  These 
are the examples that are consistent with DIS+Drell-Yan data 
sets (see text).}
\label{taugq}
\end{minipage}
\end{center}
\end{figure}

\paragraph{Uncertainty on the gluon-quark luminosity function}
In analogy with the discussion of gluon-gluon luminosities 
in the last section,  we now study the variations of gluon-quark luminosities.
Figure~\ref{taugq} shows the ratio of the  
gluon-quark luminosity normalized to the 
corresponding result from CTEQ4M, for the same variations 
used in the last section, as a function of $\sqrt{\tau}$.
Indicated on the figure is the 
region of single-top production at the Tevatron, near 
$\sqrt{\tau}=0.1$.  
The variations indicated by the solid curves do 
not include the 
uncertainty of the quark distributions,  except those 
of the sea quarks driven by different gluon variations. 
Since this is the flavor-independent sum of quarks, which 
is generally constrained to within $2-3\%$ by DIS data,  we expect
these uncertainties to be negligible.  There is one exception, 
however, and that is at very large $x$ and $Q^2$,  which is 
discussed in reference~\cite{hiquark}.  The toy model in 
that paper provides more quarks for $x>0.5$ at large $Q^2$ than 
present in CTEQ4M, and is indicated by the dotted curve 
in Fig.~\ref{taugq}.  This model is not meant to be taken 
seriously as the true set of parton distributions in 
nature,  but should be indicative of the 
current uncertainty in the quark distributions 
in this kinematic region.
The dotted curve is only significant 
at very large $\sqrt{\tau}$,  but emphasizes once again the need
for much more quality information about parton distributions
at large $x$.  

\paragraph{Summary of gluon-gluon and gluon-quark uncertainties}
We summarize in Table~\ref{tab2} what one observes in the 
last two sections.
Since the sizes of the bands were
almost identical for $\sqrt{s}=2$ or 14 TeV,  we only
give one set of numbers for the gluon-gluon uncertainty 
and one set for gluon-quark.  Above $\sqrt{\tau}>0.4$ the 
uncertainties for both gluon-gluon and gluon-quark are 
increasing rapidly and should simply be considered as being unconstrained 
with these data.  
These are not meant
to be precise uncertainties obtained by rigorous statistical
analysis,  but reasonable
error estimates in the same spirit as the estimate 
for the uncertainty in the theory that one obtains 
by varying the $\mu^2$
scale in the calculation.  They are also subjective to the
extent that they depend on the choices of what constitute
acceptable global fits. However,
we showed that even if one allows fits with total $\chi^2$
more than 200 units (for 1300 data points) greater than that of CTEQ4M,
the range
of gluon distributions for $x<0.15$ is only 5\% larger
than is shown in this table.  We therefore consider this
a robust and conservative estimate of the uncertainties
due to the parton distributions.

\begin{table}[tbph]
\begin{center}
\begin{tabular}{|c|c|c|}\hline
$\sqrt{\tau}$ Range & Gluon-Gluon & Gluon-Quark \\ \hline
$<0.1$ & $\pm$10\% & $\pm$10\% \\
0.1-0.2 & $\pm$20\% & $\pm$10\% \\
0.2-0.3 & $\pm$30\% & $\pm$15\% \\
0.3-0.4 & $\pm$60\% & $\pm$20\% \\ \hline
\end{tabular}
\end{center}
\caption{Recommended uncertainties on gluon-gluon and gluon-quark
luminosities for both the Tevatron and LHC, as a function of $\sqrt{\tau}$.
This is compared to CTEQ4M as the default parton distribution set.}
\label{tab2}
\end{table}

\paragraph{Comments on Other Data Sets}
For the analysis presented above, we have only used deep-inelastic
scattering and Drell-Yan data.  
Historically, direct photon production data were thought to place
good constraints on the gluon. However, it was pointed out some time ago
\cite{cteqdp} that the theoretical uncertainty of NLO
QCD theory was too large to allow an accurate phenomenological
analysis of direct photon data and that the available
experimental results showed a clear pattern of deviation from NLO
theory expectations. 
The recent publication of the E706 direct photon
data set~\cite{e706} has dramatically confirmed this observation
(the deviation from NLO theory is as large as a factor of 3-4), and
provided clear experimental evidence for the need to include initial state
$k_t$ effects due to multi-gluon radiation, as proposed in Ref.~\cite{cteqdp}.
Thus, theoretical progress on the resummation of multi-gluon radiation is
a prerequisite for using direct photon data as a reliable constraint on
the gluon distribution.

The original CTEQ4 analysis also included CDF and D0 single jet inclusive
data in the global fits.  Unfortunately, the experimental situation with
these data has become murkier with the more recent D0 
data analysis~\cite{d0jet}.  Even though
the CDF and D0 data sets are consistent within the 
experimental systematic error bands, they are
presently inconsistent within statistical uncertainties. Under these
circumstances, the proper treatment of the systematics has become
important. This awaits final publication of both data sets.

We note that it is unlikely that either the direct photon or high $p_t$ jet
measurements will be able to reduce significantly the uncertainties for
$x<0.15$, since the experimental normalization uncertainties
and theoretical scale dependence is typically 10\%.
The main contribution these processes can give
in the future is at large $x$, and in the shape of the gluon distribution
at medium values of $x$.

Our baseline variations of the gluon distribution
can be used as the benchmark for comparison with present 
and future measurements which
have the potential to narrow the uncertainties.
For this purpose the various parton distribution sets
which typify the variations shown in this paper will be made available
to interested users~\cite{web}.

\paragraph{Conclusions}
We have studied the uncertainty in the gluon
distribution of the nucleon by systematically 
varying the relevant parameters in the 
QCD global analysis.  This uncertainty dominates 
the current uncertainty in the calculation of 
many important new processes at the Tevatron 
and LHC.
The uncertainty 
depends critically on parton $x$ and $Q^2$.
We present a table of estimated uncertainties 
for integrated gluon-gluon and gluon-quark luminosities for both
the Tevatron and LHC as a function of 
$\sqrt{\tau}=\sqrt{x_1 x_2}=\sqrt{\hat{s} /s}$.
The uncertainties are reasonably small, except for 
large $x$,  where future emphasis should be 
placed.

\begin{thebibliography}{99}

\bibitem{cteq4}  H.L. Lai et al., Phys. Rev. D55:1280, (1997).

\bibitem{alekhin} S. Alekhin, hep-ph/9611213, submitted to 
Elsevier Science

\bibitem{mrs} A.D. Martin et al., Phys. Lett. B387:419, (1996).

\bibitem{dis97} Proceedings of the 5th International Workshop 
on Deep Inelastic Scattering and QCD, DIS97, AIP Conference 
Proceedings No. 407.

\bibitem{alpha} Particle Data Group, Phys. Rev. D54:1, (1996)

\bibitem{ehlq} E. Eichten et al., Rev. Mod. Phys. 56:579, (1984)

\bibitem{hiquark} S. Kuhlmann et al., Phys. Lett. B409:271, (1997).

\bibitem{cteqdp} J. Huston et al., Phys. Rev D51:6139, (1995)

\bibitem{e706} E706 Collaboration (L. Apanasevich et al.),  
hep-ex/9711017, FNAL-Pub-97/351-E

\bibitem{d0jet} D0 Collaboration, ``The Dijet Mass Spectrum at D0'', 
{\it International Europhysics Conference on High Energy Physics,}
August 19-26, 1997, Jerusalem, Israel

\bibitem{web} All of the eight parton distribution variations 
described in this paper are available from the CTEQ web site, 
http://www.phys.psu.edu/$\sim$cteq/.
 
\end{thebibliography}
\end{document}